\documentclass{osa-article}
 \usepackage{tensor}
 \usepackage{physics}
\definecolor{authorcolor}{RGB}{148,54,52}
\definecolor{urlblue}{RGB}{46,46,177}
\def\@journalname{Optics Express}
\articletype{Research Article}

\begin{document}

\title{Using all transverse degrees of freedom in quantum communications based on a generic mode sorter}

\author{Yiyu Zhou,\authormark{1,9,*} Mohammad Mirhosseini,\authormark{1,2,9} Stone Oliver,\authormark{1,3} Jiapeng Zhao,\authormark{1} Seyed Mohammad Hashemi Rafsanjani,\authormark{1,4} Martin P. J. Lavery,\authormark{5} Alan E. Willner,\authormark{6} and Robert W. Boyd\authormark{1,7,8}}

\address{\authormark{1}The Institute of Optics, University of Rochester, Rochester, New York 14627, USA\\
\authormark{2}Thomas J. Watson, Sr., Laboratory of Applied Physics, California Institute of Technology, Pasadena, California 91125, USA\\
\authormark{3}Department of Physics, Miami University, Oxford, Ohio 45056, USA\\
\authormark{4}Department of Physics, University of Miami, Coral Gables, Florida 33146, USA\\
\authormark{5}School of Engineering, University of Glasgow, Glasgow, Scotland G12 8LT, UK\\
\authormark{6}Department of Electrical Engineering, University of Southern California, Los Angeles, California 90089, USA\\
\authormark{7}Department of Physics, University of Ottawa, Ottawa, Ontario K1N 6N5, Canada\\
\authormark{8}boyd@optics.rochester.edu\\
\authormark{9}These authors contributed equally
}

\email{\authormark{*}yzhou62@ur.rochester.edu} 



\begin{abstract}
The dimension of the state space for information encoding offered by the transverse structure of light is usually limited by the finite size of apertures. The widely used orbital angular momentum (OAM) number of Laguerre-Gaussian (LG) modes in free-space communications cannot achieve the theoretical maximum transmission capacity unless the radial degree of freedom is multiplexed into the protocol. While the methodology to sort the radial quantum number has been developed, the application of radial modes in quantum communications requires an additional ability to efficiently measure the superposition of LG modes in the mutually unbiased basis. Here we develop and implement a generic mode sorter that is capable of sorting the superposition of LG modes through the use of a mode converter. As a consequence, we demonstrate an 8-dimensional quantum key distribution experiment involving all three transverse degrees of freedom: spin, azimuthal, and radial quantum numbers of photons. Our protocol presents an important step towards the goal of reaching the capacity limit of a free-space link and can be useful to other applications that involve spatial modes of photons.
\end{abstract}

\section{Introduction}

{Within the past few decades the orbital angular momentum (OAM) modes have received extensive attention \cite{mair2001entanglement,karimi2014generating,lavery2017free,bozinovic2013terabit,fickler2012quantum} and are widely applied in various information technologies, including quantum teleportation \cite{wang2015quantum}, optical communications \cite{wang2012terabit}, and quantum key distribution \cite{mirhosseini2015high,sit2017high,zhao2018performance}. Compared to the intrinsically bounded polarization state space, the OAM modes offer an infinite-dimensional Hilbert space for information encoding and therefore can be used to increase the transmission rate of a communication link. However, the finite size of apertures in a realistic system usually constrains the dimension of OAM state space that can be accessed. On the other hand, the OAM modes only account for azimuthal variations in the transverse plane, and it has been shown that these modes cannot reach the capacity limit of a communications link without including the radial degree of freedom \cite{zhao2015capacity}. It is thus highly desirable to multiplex the remaining radial degree of freedom to increase the transmission rate for both classical and quantum communications.}

{A complete and orthonormal basis that incorporates both radial and azimuthal variations can be constituted by Laguerre-Gaussian (LG) modes. The LG modes are characterized by a radial quantum number $p$ and an OAM quantum number $\ell$, and arbitrary paraxial field can be described by these modes \cite{allen1992orbital}. Notably, LG modes have been shown to be good approximations to the eigen propagation modes of circular apertures \cite{rodenburg2015communicating}, and their small divergence angle and intrinsic rotational symmetry make these modes preferable in free-space optical communications. Moreover, it is apparent that the transmission rate of a communications system can be further increased by using both azimuthal and radial degrees of freedom, and it has been shown that the radial index $p$ can potentially mitigate the power loss when the receiver has a limited aperture size \cite{li2017power}. In addition, different from the vortex phase structure related to the OAM index, the radial index corresponds to a radial, amplitude-only distribution. While the phase structure can be strongly distorted by atmospheric turbulence \cite{boyd2011influence}, the intensity pattern of the transmitted beam can remain recognizable for a 1.6 km free-space link \cite{lavery2017free}, and an intensity pattern recognition accuracy can be higher than 98\% for a 3 km link \cite{krenn2014communication} and 80\% for a 143 km link \cite{krenn2016twisted}, which suggests that the amplitude structure associated with a nonzero radial index may be helpful to turbulence mitigation \cite{wang2018Towards}. Recent advances have shown how to efficiently measure and use $p$ and $\ell$ in optical communications \cite{yiyu2017sorting,berkhout2010efficient,mirhosseini2013efficient, larocque2017generalized,gu2018gouy,fu2018realization,trichili2016optical,mafu2013higher}, but the mode sorting in mutually unbiased basis of LG modes required by a quantum key distribution (QKD) protocol has not previously been reported. While a deterministic detection scheme for mutually unbiased bases of polarization and OAM degrees of freedom has been developed \cite{ndagano2017deterministic}, the same strategy cannot be directly applied to spatial modes due to the lack of efficient manipulation devices such as half-wave plates for spatial modes. Besides the usefulness in QKD, the capability of sorting superposition modes can be helpful to other quantum applications such as certifying high-dimensional entanglement \cite{bavaresco2018measurements} as well as superresolution imaging \cite{tsang2017subdiffraction}. In the following we present how to build a superposition mode sorter and provide the experimental demonstration of a QKD protocol employing all three transverse degrees of freedom.}

\section{Construction of mutually unbiased bases}
To develop a BB84 protocol one needs at least two sets of bases that are mutually unbiased: Every element in each basis is a uniform superposition of elements in another basis, which guarantees that measurement in the wrong basis reveals no information of the measured state. Here we first examine each transverse degree of freedom individually and then demonstrate how to construct two mutually unbiased bases for QKD. The first degree of freedom employed in QKD protocol is the polarization of photons \cite{bennett2014quantum}, which provides a two-dimensional state space spanned by horizontal $\ket{{H}}$ and vertical $\ket{{V}}$ polarizations. While polarization is the most widely used degree of freedom in QKD, the azimuthal structures of light have recently seen increasing usage due to their direct relation with the OAM of photons \cite{mirhosseini2015high}. Each photon can carry an OAM of $l\hbar$ \cite{allen1992orbital} and the corresponding OAM state can be written as $\ket{l}_{\ell}$, where $l$ is an arbitrary integer and the subscript $\ell$ denotes an OAM state. In this work we restrict ourselves to an OAM sub-space that is spanned by $\ket{\pm 2}_{\ell}$. Finally we use the radial quantum number $p$ of LG functions as another independent resource for information encoding \cite{pang2018demonstration,karimi2014exploring}. Again we restrict ourselves to a sub-space that is spanned by the two lowest radial indices $p =0, 1$. The direct product of these three sub-spaces constitutes an 8-dimensional Hilbert space that is used as our first basis for QKD. The elements in this basis are $\{\ket{H,0_p,-2_\ell}, \ket{H,0_p,2_\ell}, \cdots, \ket{V,1_p,2_\ell} \}$. We refer to this basis as the \emph{main} basis. The above-specified choices of bases for azimuthal and radial degrees of freedom allows us to employ recent advances in sorting LG modes according to both OAM and radial indices \cite{yiyu2017sorting,berkhout2010efficient,mirhosseini2013efficient,larocque2017generalized}.

\begin{figure}[t]
\centering
\includegraphics[width=0.6  \linewidth]{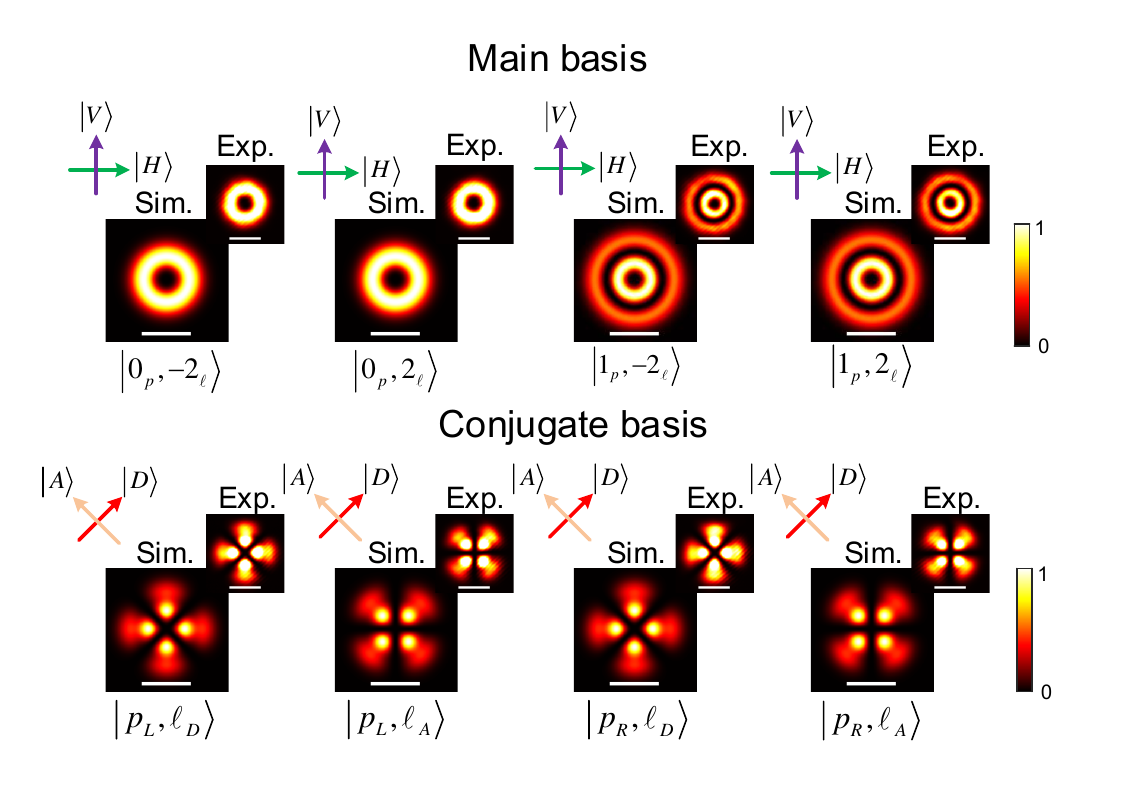}
\caption{Spatial representations of elements in the main and conjugate bases. The normalized intensity distribution of each spatial mode is presented. The upper right image is the corresponding experimental (Exp.) record of the simulated (Sim.) image for comparison. Each mode can have the two different polarizations as shown at the upper left corner. Scale bar, 1 mm.}
\label{fig:SpatialMode}
\end{figure}

To implement secure QKD we need an additional capability of measuring photons in at least one conjugate basis. The conjugate basis is mutually unbiased with respect to the main basis, and can be constructed through a direct product of respective complementary bases of polarization, azimuthal, and radial sub-spaces \cite{nagali2010experimental}. The complementary basis of polarization sub-space can be built as
\begin{equation}\label{Eq:DAdefinition}
\ket{D} = \frac{\ket{H} +  \ket{V} }{\sqrt{2}},\hspace{1cm} \quad \ket{A} = \frac{\ket{H} -  \ket{V} }{\sqrt{2}},
\end{equation}
where $\ket{D}$ and $\ket{A}$ denote the diagonal and anti-diagonal polarizations respectively. We use a similar choice for OAM states and define
\begin{equation}\label{Eq:OAMdefinition}
 \begin{aligned}
\ket{\ell_D} = \frac{\ket{ -2}_{\ell}+\ket{2}_{\ell}}{\sqrt{2}}, \hspace{1cm}
 \ket{\ell_A} = \frac{\ket{ -2}_{\ell}-\ket{2}_{\ell}}{\sqrt{2}},
 \end{aligned}
\end{equation}
The complementary basis for radial modes is taken to be
 \begin{equation}\label{Eq:pdefinition}
 \begin{aligned}
\ket{p_L} = \frac{ \ket{0}_{p}+i \ket{1}_{p}}{\sqrt{2}}, \hspace{1cm}
 \ket{p_R} = \frac{ \ket{0}_{p}-i \ket{1}_{p}}{\sqrt{2}},
 \end{aligned}
\end{equation}
where the subscript $L$ and $R$ follow the notation of left- and right-handed circular polarization. We choose this definition because such states are easier to generate experimentally, but we stress that this will not make any fundamental change to the QKD protocol. With all these definitions, the elements in the complementary basis, referred to as \emph{conjugate} basis, are \{$\ket{D, p_L, \ell_D}$, $\ket{D, p_{L}, \ell_{A}}$, $\cdots$, $\ket{A, p_{R}, \ell_{A}} $\}. Spatial intensity distributions of all these modes are given in Fig.~\ref{fig:SpatialMode}.

\section{Construction of a superposition mode sorter}

Having established the two mutually unbiased bases, we need to perform coherent detection in each of these bases. For the main basis, devising a coherent detection strategy is straightforward. A sequence of a polarizing beamsplitter (PBS), a radial mode sorter \cite{yiyu2017sorting}, and an OAM sorter \cite{berkhout2010efficient} can losslessly project input photons onto elements in the main basis. However, each state in the conjugate basis is a superposition of different LG modes and cannot be efficiently measured by using radial and OAM sorter only. To address this problem, we develop a generic, scalable scheme for sorting such superposition states and experimentally realize it as a part of the QKD protocol. For simplicity we only focus on the radial degree of freedom thereafter, and the same scheme can be directly applied to other degrees of freedom.

\begin{figure}[!t]
\centering
\includegraphics[width= 0.8 \linewidth]{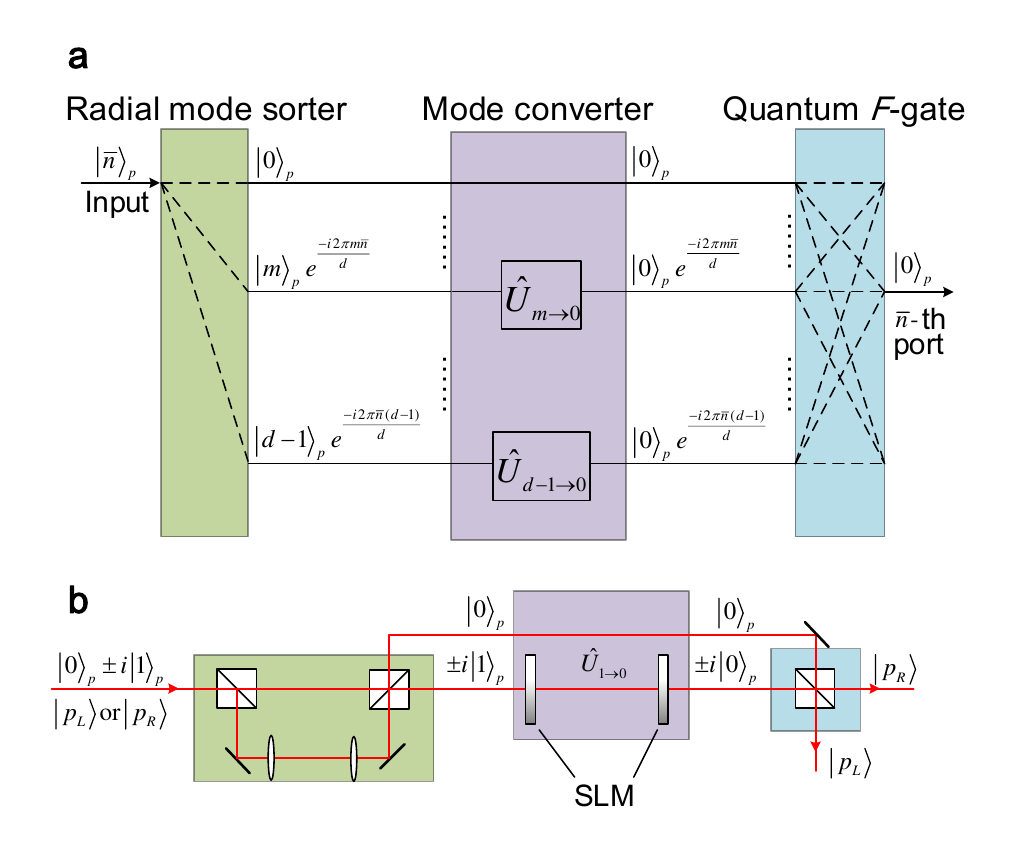}
\caption{Schematic of the radial superposition mode sorter. (a) Conceptual schematic of a $d$-dimensional radial superposition mode sorter. (b) Two-dimensional ($d=2$) realization to sort $\ket{p_L}$ and $\ket{p_R}$. The unitary transformation $\hat{U}_{1\rightarrow 0}$ is realized by SLMs. The modes $\ket{p_L}$ and $\ket{p_R}$ are directed to different output ports as indicated at the last beamsplitter.}
\label{fig:MUBsorter}
\end{figure}

A conceptual schematic for a radial superposition mode sorter is shown in Fig.~\ref{fig:MUBsorter}. For a $d$-dimensional Hilbert space spanned by radial modes $\ket{m}_p$, where $m \in \{0, 1, \cdots, d-1\}$, the corresponding complementary basis can be defined as
\begin{equation}
\begin{aligned}
\ket{\bar{n}}_p = \frac{1}{\sqrt{d}}\sum_{m=0}^{d-1}\exp \left(  \frac{-i2\pi m\bar{n}}{d}  \right) \ket{m}_p,
\label{eq:MUBdef}
\end{aligned}
\end{equation}
where $\bar{n} \in \{0, 1, \cdots, d-1\}$ is the superposition mode index, and a bar above the number indicates that this is an element in the complementary basis. The port of each state is labelled by another distinct ket $\ket{k}$, where $k \in \{0, 1, \cdots, d-1\}$. Initially, all superposition modes are located at the same port, which can be expressed as $\ket{\bar{n}}_p  \ket{0}$. We first apply a radial mode sorter to direct different radial mode components towards distinct, non-overlapping ports \cite{yiyu2017sorting}, which can be expressed as
\begin{equation}
\begin{aligned}
  \ket{\bar{n}}_p\ket{0}  &=   \frac{1}{\sqrt{d}}\sum_{m=0}^{d-1}\exp \left(  \frac{-i2\pi m\bar{n}}{d}  \right) \ket{m}_p  \ket{0}   \rightarrow  \frac{1}{\sqrt{d}}\sum_{m=0}^{d-1}\exp \left(  \frac{-i2\pi m\bar{n}}{d}  \right) \ket{m}_p    \ket{m}.
\end{aligned}
\end{equation}
Then a mode converter performs a unitary transformation $\hat{U}_{m \rightarrow 0}$ to convert individual radial modes $\ket{m}_p$ to the same state $\ket{0}_p$, which enables effective interference between these otherwise orthogonal modes and the consequent state becomes
\begin{equation}
\begin{aligned}
\quad  \frac{1}{\sqrt{d}}\sum_{m=0}^{d-1}\exp \left(  \frac{-i2\pi m\bar{n}}{d}  \right) \ket{m}_p    \ket{m} \rightarrow  \frac{1}{\sqrt{d}}\sum_{m=0}^{d-1}\exp \left(  \frac{-i2\pi m\bar{n}}{d}  \right) \ket{0}_p \ket{m}.
\end{aligned}
\end{equation}
Finally a discrete Fourier transform performed by a quantum $F$-gate can direct photons to different output port indexed by the superposition mode index $\bar{n}$ as \cite{ionicioiu2016sorting}
\begin{equation}
\begin{aligned}
\quad & \hat{F} \Big[      \frac{1}{\sqrt{d}} \sum_{m=0}^{d-1}  \exp \left(  \frac{-i2\pi m\bar{n}}{d}  \right)\ket{0}_p \ket{m} \Big]  =  \frac{1}{\sqrt{d}}\sum_{m=0}^{d-1}\exp \left(  \frac{-i2\pi m\bar{n}}{d}  \right)  \hat{F}\Big[ \ket{0}_p \ket{m} \Big]\\
&= \frac{1}{d}\sum_{j=0}^{d-1}  \sum_{m=0}^{d-1} \exp \left(  \frac{i2\pi m(j-\bar{n})}{d}  \right) \ket{0}_p  \ket{j}= \sum_{j=0}^{d-1} \delta (j-\bar{n}) \ket{0}_p  \ket{j} = \ket{0}_p  \ket{\bar{n}},
\label{eq:mubsorter}
\end{aligned}
\end{equation}
where the operation of quantum $F$-gate is defined as
\begin{equation}
\begin{aligned}
\hat{F} \Big[ \ket{m}_p \ket{k}  \Big]=\frac{1}{\sqrt{d}}\sum_{j=0}^{d-1} \text{exp}\left(\frac{i2\pi jk}{d}\right)\ket{m}_p  \ket{j}.
\end{aligned}
\end{equation}
Therefore, through the use of a mode converter, the superposition mode $\ket{\bar{n}}_p $ can be efficiently sorted to $\bar{n}$-th output port with in principle unity efficiency and zero cross-talk.

\begin{figure}[!t]
\centering
\includegraphics[width=  \linewidth]{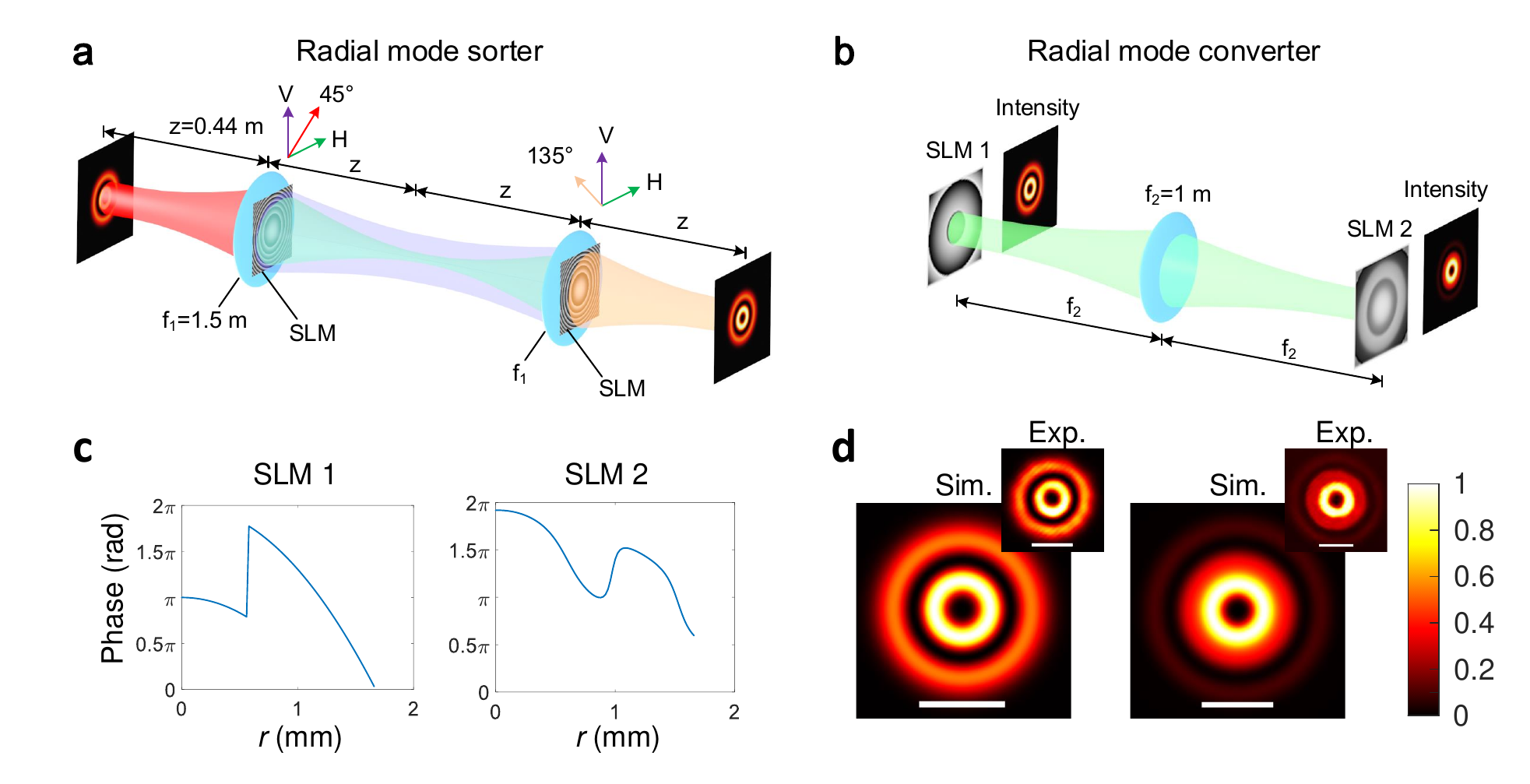}
\caption{(a) Common-path radial mode sorter. (b) Radial mode converter which transforms $\ket{1_p,2_{\ell}}$ to $\ket{0_p,2_{\ell}}$. (c) The phase distribution as a function of radius $r$ on the two SLMs in the radial mode converter. (d) The input and output spatial mode intensity distribution of the radial mode converter. The upper right image is the corresponding experimental (Exp.) record of the simulated (Sim.) image for comparison.}
\label{fig:MUBsorterExp}
\end{figure}

A two-dimensional realization of this scheme to sort $\ket{p_L}$ and $\ket{p_R}$ is shown in Fig.~\ref{fig:MUBsorter}(b) as an intuitive example, where the binary radial mode sorter follows the scheme in \cite{yiyu2017sorting} and is realized by a Mach-Zenhder interferometer with extra lenses in one arm to perform the Fourier transform. After the radial mode sorter, the radial modes $\ket{0}_p$ and $\ket{1}_p$ are sorted to different paths, but the relative phase determined by the definition of $\ket{p_L}$ and $\ket{p_R}$ as in Eq.~(\ref{Eq:pdefinition}) persists. The relative phase is represented by the factor $\pm i$ in Fig.~\ref{fig:MUBsorter}(b). The subsequent mode converter realized by spatial light modulator (SLM) transforms the LG mode $\ket{1}_p$ to $\ket{0}_p$ \cite{morizur2010programmable,Saumya2018Measurement}, and the phase imparted to the SLMs can be calculated by a nonlinear fitting algorithm \cite{wong2016high}. Then a beamsplitter recombines the modes and acts as a binary quantum $F$-gate \cite{ionicioiu2016sorting}. Before the final beamsplitter, the incident modes at the input ports can be written as $\ket{0}_p$ and $\pm i \ket{0}_p$ respectively, where the factor $\pm i$ represents the relative phase between the initial $\ket{0}_p$ and $\ket{1}_p$ states. Due to the interference at the beamsplitter, the sign of the phase term $\pm i$ determines the port that the photon will be directed into. Therefore, through the use of a radial mode sorter, a mode converter realized by the SLMs, and a binary quantum $F$-gate realized by a beamsplitter, $\ket{p_L}$ and $\ket{p_R}$ can be efficiently separated to different output ports. We note that this scheme should also be applicable to other spatial modes such as Hermite-Gaussian (HG) modes given the existence of a corresponding mode sorter \cite{linares2017spatial} and converter \cite{morizur2010programmable}, which can be useful to realize super-resolution imaging \cite{tsang2017subdiffraction}.

\begin{figure}[!t]
\centering
\includegraphics[width=  0.6 \linewidth]{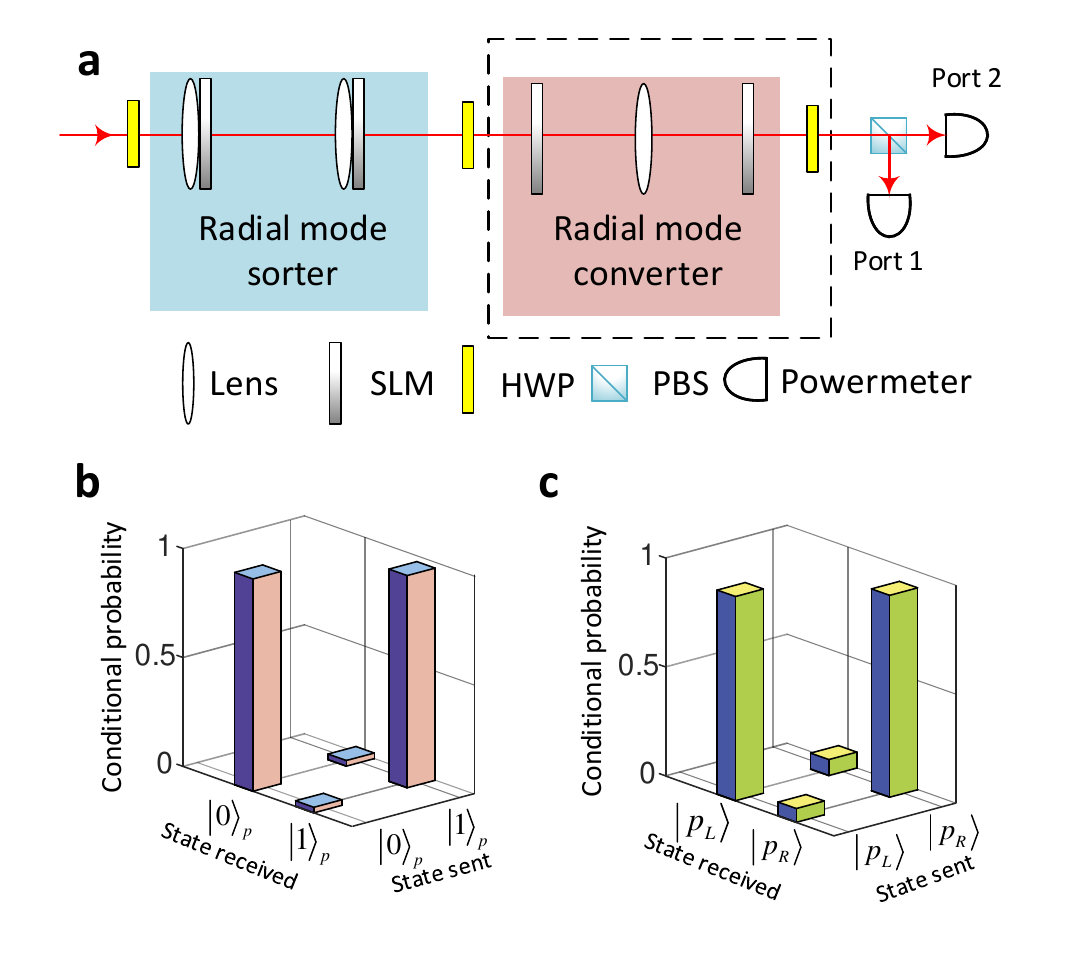}
\caption{(a) Experimental realization of a common-path radial superposition mode sorter. By removing the radial mode converter in the dashed box it becomes a common-path radial mode sorter. (b) The measured cross-talk of the common-path radial mode sorter. (c) The measured cross-talk of the common-path radial superposition mode sorter.}
\label{fig:classicalcross}
\end{figure}

In our experiment, we use polarization-sensitive SLMs to develop a stabilized, common-path radial mode sorter and converter as shown in Figs.~\ref{fig:MUBsorterExp}(a) and \ref{fig:MUBsorterExp}(b), respectively. The common-path radial mode sorter is composed of two polarization-sensitive SLMs, each with a spherical lenses attached respectively. Each spherical lens is imaged onto the respective SLM by a 4-$f$ system in the experiment. A quadratic phase pattern which is equivalent to a lens of focal length 0.62~m is imprinted on both SLMs. Both spherical lenses have a focal length of $f_1=1.5$~m. The distance between the injected mode beam waist plane and the first SLM is $z= 0.44 \text{ m}$, and the separation between two SLMs is $2z$. The output plane of the radial mode sorter is of distance $z$ after the last SLM. The Fourier transforming lens in radial mode converter has a focal length of 1 m. In this setup two polarizations are employed as two arms of a Mach-Zenhder interferometer, and the injected mode is 45 degree polarized. Since vertically polarized light is not affected by SLMs, one can check that two spherical lenses perform a Fourier transform to it \cite{yiyu2017sorting}. Horizontal polarization, however, is modulated by both SLMs. Each SLM with the attached lens becomes a lens of focal length 0.44~m, exactly equal to the propagation distance $z$. Hence, horizontally polarized light experiences two consecutive Fourier transforms. If we treat vertically polarized beam as the reference arm, then horizontally polarized beam will gain a mode-dependent Gouy phase of $\exp(-ip\pi)$ due to its extra Fourier transform. One can check that this phase is $0$ for $\ket{0}_p$ and $-\pi$ for $\ket{1}_p$ \cite{yiyu2017sorting}, and thus $\ket{0}_p$ remains 45 degree polarized, but the polarization of $\ket{1}_p$ is rotated to 135 degrees. In the experiment we use a half-wave plate (HWP) so that $\ket{0}_p$ is vertically polarized while $\ket{1}_p$ becomes horizontally polarized at the output port. Compared to the sorter based on Mach-Zenhder interferometer, this common-path interferometer is robust to vibration and air turbulence. The beam waist radius of LG modes used in our experiment is $w_0=462.3$~$\mu$m, and we note that the parameters of the radial mode sorter mentioned above are specific to this beam waist radius and cannot be directly applied to LG modes of different beam size \cite{yiyu2017sorting}. However, one can always use a 4-$f$ system with appropriate lateral magnification to adjust the beam waist radius. In addition, the OAM index $\ell$ of the incident LG mode can also affect the radial mode sorter because the Gouy phase is also a function of $|\ell|$. In this experiment we use $\ell=-2$ and $\ell=2$, so the value of $|\ell|$ is a constant and thus its effect can be ignored. To remove the $\ell$-dependence of radial mode sorter, one can use a Dove prism to cancel the $\ell$-dependent phase as demonstrated in \cite{fu2018realization}.

The common-path radial mode converter is implemented by two SLMs connected by a Fourier transforming lens to realize mode conversion from $\ket{1_p,\pm 2_{\ell}}$ to $\ket{0_p,\pm 2_{\ell}}$. Taking into account the fact that the converter only reshapes the radial structure of the mode, the phase written on SLMs should be a function of radius $r$ only. So the phase on first SLM can be decomposed to the polynomials of radius and we use nonlinear fitting algorithm \cite{wong2016high} to improve the conversion efficiency by adjusting coefficients of these polynomials. The second SLM is used to cancel the residue phase of the converted mode because $\ket{0_p,\pm 2_{\ell}}$ possesses a flat phase structure in the radial direction. We choose three polynomials \{$r^2$,  $r^3$, $r^4$\} in the algorithm. We tested that more polynomials (up to $r^7$) can provide negligible conversion efficiency improvement while costing much more time to run the code. The phase distributions on two SLMs are shown in Fig.~\ref{fig:MUBsorterExp}(c). The intensity distributions of the input and output fields are shown in Fig.~\ref{fig:MUBsorterExp}(d). It can be seen that the converted mode is similar but not identical to $\ket{0}_p$ due to the non-unity conversion efficiency. The conversion efficiency in our simulation is $|\braket{0_p}{0'_p}|^2=82.7\%$ where $\ket{0'_p}$ denotes the converted mode by SLMs. Due to this non-unity conversion efficiency, the minimum cross-talk of the superposition mode converter can be readily calculated to be 4.5\%. In general, a mode conversion requires multi-plane iterations \cite{labroille2014efficient}, and therefore our implementation to transform $\ket{1}_p$ to $\ket{0}_p$ is not sufficient to achieve a unity conversion efficiency. However, it is straightforward to cascade more SLMs to reduce the cross-talk of the superposition mode sorter \cite{morizur2010programmable}.

The radial superposition mode sorter we experimentally implemented is depicted in Fig.~\ref{fig:classicalcross}(a), and its performance is evaluated by measuring the cross-talk matrix. We first remove the radial mode converter such that the set-up becomes a common-path radial mode sorter. We characterize the performance of this common-path radial mode sorter by measuring output power of two output ports of the PBS and the result is presented in Fig.~\ref{fig:classicalcross}(b). The cross-talk of the radial mode sorter is around 2.6\%, which is defined as the power in the wrong port divided by the total output power when a radial mode is injected. We note that this sorter can also be applied to HG modes because both LG modes and HG modes are the eigenmodes of the fractional Fourier transform. We then cascade the radial mode converter and inject $\ket{p_L}$ and $\ket{p_R}$ mode to the superposition mode sorter. The measured cross-talk is shown in Fig.~\ref{fig:classicalcross}(c), which is around 7.4\%. As mentioned above, the cross-talk of the superposition mode sorter can in principle be further decreased by cascading mode SLMs to perform the mode conversion.

 \section{Implementation of QKD}

\begin{figure*}[!t]
\centering
\includegraphics[width=  \textwidth]{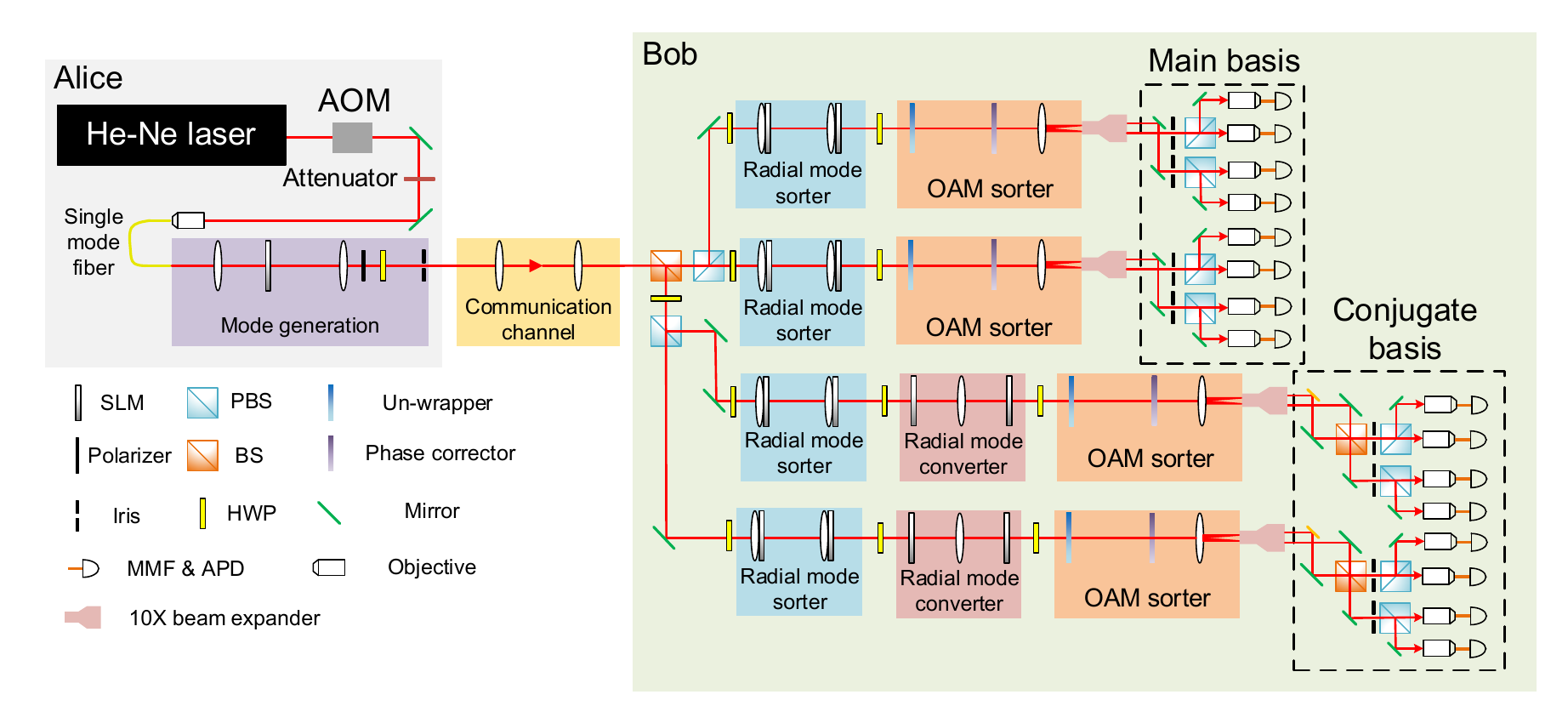}
\caption{Schematic of the QKD protocol involving three degrees of freedom. Relay lenses are omitted for simplicity. }
\label{fig:setup}
\end{figure*}

Having constructed the sorters for both mutually unbiased bases, we next demonstrate the implementation of an 8-dimensional QKD protocol involving all three degrees of freedom. A schematic diagram of the setup is presented in Fig.~\ref{fig:setup}. A He-Ne laser is modulated by an acousto-optic modulator (AOM) to generate 200 ns optical pulses. The average photon number in each pulse is attenuated by neutral density filters and crossed polarizers to $\mu=0.1$. Computer-generated holograms imparted on SLM are used to generate LG modes of beam waist size $w_0=462.3$~$\mu$m at the first diffraction order \cite{mirhosseini2013rapid}. A combination of PBS and HWP is used to measure the polarization states in both mutually unbiased bases. Then an OAM mode sorter consisting of an un-wrapper and a phase corrector \cite{lavery2012refractive} is used to sort the two OAM modes $\ket{-2}_{\ell}$ and $\ket{2}_{\ell}$ \cite{berkhout2010efficient}. An additional beamsplitter (BS) is needed to recombine the separated modes so as to sort OAM superposition states $\ket{\ell_D}$ and $\ket{\ell_A}$. We note that the sorting mechanism of OAM superposition states is the same as that of the superposition of radial modes. However, the mode converter is not needed because the log-polar transformation induced by the OAM mode sorter already converts the OAM modes to have the same spatial shape, and therefore simply injecting the sorted OAM modes to a beamsplitter ($i.e.$, a two-dimensional quantum $F$-gate) can efficiently separate the OAM superposition states. The sorting mechanism for the radial degree of freedom follows the scheme presented in Fig.~\ref{fig:MUBsorter}. After the polarization state is determined, the common-path radial mode sorter is be used to map different radial modes to different polarizations, therefore one can use a PBS to detect the radial quantum number. A radial mode converter needs to be inserted to sort the radial superposition modes $\ket{p_L}$ and $\ket{p_R}$ as discussed above. All sorted photons are collected by multi-mode fibers (MMFs) and detected by single-photon avalanche photodiodes (APDs, Perkin Elmer SPCM-AQRH-14-FC). We note that since only four APDs are available at the time of performing experiment, at Bob's side the data is collected for elements in each basis separately and combined later. Since the output of the OAM mode sorter has a small size, we use a 10X beam expander (GBE10-A, Thorlabs) to expand the beam before the MMFs.

\begin{figure}[t]
\centering
\includegraphics[width= 0.8\linewidth]{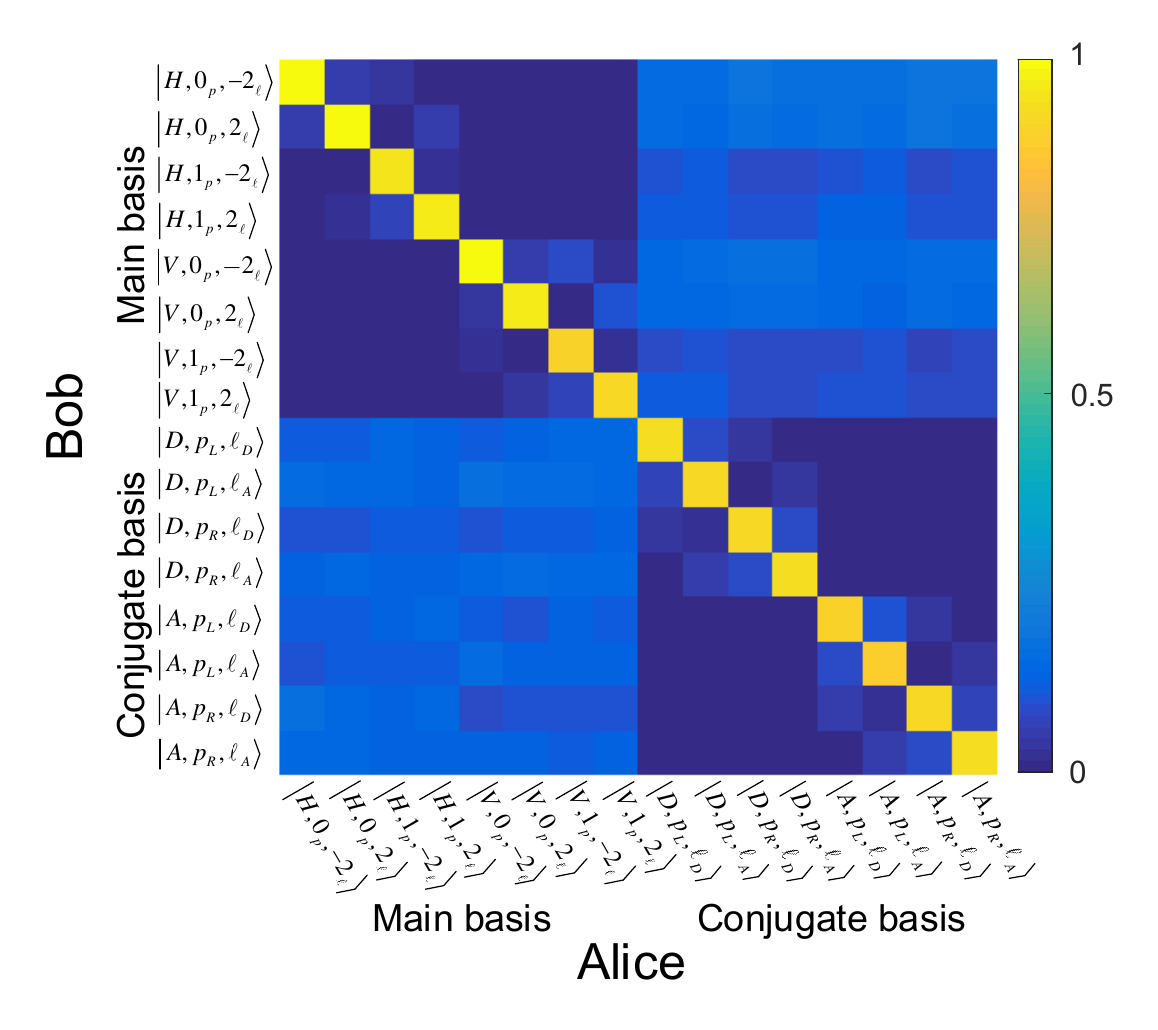}
\caption{Conditional probability matrix measured with $\mu = 0.1$ pulses. Each element in this matrix represents the probability that Bob detects the corresponding symbol conditioned on the symbol sent by Alice.  }
\label{fig:crosstalk}
\end{figure}

To evaluate the performance of the superposition mode sorter, we measure the cross-talk matrix with highly attenuated coherent states and the result is presented in Fig.~\ref{fig:crosstalk}. The cross-talk ranges from 6.0\% to 16.7\%, with an average of 11.7\% well below the 8-D QKD error threshold 24.7\% \cite{cerf2002security}. Here the cross-talk is the probability that the photon triggers the wrong APDs when Bob is detecting in the correct basis. The mutual information is calculated to be $I_{\text{AB}}=2.15$ bits per sifted photon (see Appendix A). We also note that the SLM used in our experiment can be readily replaced by passive, polarization-dependent liquid crystal retarder to realize a scalable, low-cost sorter.

{Compared to a QKD protocol with OAM encoding only \cite{mirhosseini2015high}, our protocol employs the slowly divergent LG modes and are thus practical for free-space links with finite-sized apertures \cite{rodenburg2015communicating}. In a realistic free-space channel, atmospheric turbulence can lead to modal cross-talk and reduce the transmission rate. A recent experiment \cite{lavery2017free} has suggested the potential of LG modes in a free-space channel in an urban environment. The intensity pattern of the transmitted beam can remain recognizable after a 1.6 km free-space link, and thus adaptive optics can be potentially used to mitigate turbulence \cite{lavery2017free}, which can be subject to future study. In a realistic free-space link, in addition to the spatial distortion induced by turbulence which can be corrected by adaptive optics, different LG modes accumulate different amount of Gouy phase which can affect the sorting of superposition modes \cite{zhao2018performance}. The Gouy phase for a LG mode can be written as $\phi=(2p+|\ell|+1) \arctan (z/z_R)$, where $z_R$ is the Rayleigh range and $z$ is the propagation distance. As proposed in \cite{zhao2018performance}, pre-compensation can be used in mode preparation at the transmitter's side to guarantee that the mode-dependent phase is cancelled at the receiver's side. Furthermore, the phase-dependent phase can also be removed at the receiver's side. Here we take the scheme in Fig.~\ref{fig:MUBsorter}(a) as an example to show how this mode-dependent phase can be removed. Since a radial mode sorter is used in the superposition mode sorter, individual radial modes are separated to different paths. By simply adjusting the path lengths for each radial mode, a mode-dependent phase can be added to cancel the Gouy phase. In our experiment, the radial modes are sorted to horizontal and vertical polarizations respectively, and by adding a constant phase on the subsequent polarization-sensitive SLM, we are able to compensate the relative phase between different radial modes. And the same method can be applied to the OAM index. On the other hand, the large communication bandwidth offered by LG modes \cite{lavery2013efficient} can be obtained with less difficulty for a smaller distance and thus provides benefits to short-range optical interconnects~\cite{yu2015potentials}.}

\section{Conclusion}

In conclusion, we provide an experimental demonstration of a QKD protocol which encodes information using all possible transverse degrees of freedom, i.e. polarization, radial, and OAM modes, with a resulting transmission of 2.15 bits per sifted photon. A sorting scheme for superposition spatial modes is implemented to enable this 8-D protocol and can find direct application in other fields such as super-resolution imaging and high-dimensional entanglement certification. We believe our demonstration opens up a way to fully exhaust the information resources of finite-sized apertures and therefore reach the capacity limit of a communication channel. The slowly divergent LG beams also make this protocol promising for a free-space communication network.

\section*{Appendix}
\subsection*{A. Calculation of mutual information}
For our protocol, the two bases are mutually unbiased to each other, so Alice would send each symbols with equal probability as the BB84 protocol. Here we assume a uniform error rate for detecting each mode, so the mutual information can be expressed as \cite{cerf2002security}
\begin{equation}
\begin{aligned}
I_{\text{AB}}=\log _2 d + F\log _2 F +(1-F)\log _2 \left( \frac{1-F}{d-1} \right),
\label{eq:MutualInfo}
\end{aligned}
\end{equation}
where $d=8$, the average error rate $\delta =11.7\%$, and $F$ is the probability of correct measurement $F=1-\delta=88.3\%$. With these numbers we can immediately get $I_{\text{AB}}=2.15$ bits per sifted photon.

\section*{Funding}
U.S. Office of Naval Research (grant No. N000141712443 and N000141512635); Canada Excellence Research Chairs Program; Natural Science and Engineering Research Council of Canada; Canada First Research Excellence Fund award in Transformative Quantum Technologies; National Science Foundation (No. PHY-1460352).

\end{document}